# Scalable fiber integrated source for higher-dimensional path-entangled photonic quNits

Christoph Schaeff,[1,2,*] Robert Polster,[1] Radek Lapkiewicz,[1,2] Robert Fickler,[1] Sven Ramelow,[1,2] and Anton Zeilinger[1,2,3]

[1]*Quantum Optics, Quantum Nanophysics, Quantum Information, University of Vienna, Boltzmanngasse 5, Vienna A-1090, Austria*
[2]*Institute for Quantum Optics and Quantum Information, Boltzmanngasse 3, Vienna A-1090, Austria*
[3]*Vienna Center for Quantum Science and Technology, Faculty of Physics, University of Vienna, Boltzmanngasse 5, Vienna A-1090, Austria*
*[*]Christoph.Schaeff@univie.ac.at*

**Abstract:** Integrated photonic circuits offer the possibility for complex quantum optical experiments in higher-dimensional photonic systems. However, the advantages of integration and scalability can only be fully utilized with the availability of a source for higher-dimensional entangled photons. Here, a novel fiber integrated source for path-entangled photons in the telecom band at 1.55μm using only standard fiber technology is presented. Due to the special design the source shows good scalability towards higher-dimensional entangled photonic states (quNits), while path entanglement offers direct compatibility with on-chip path encoding. We present an experimental realization of a path-entangled two-qubit source. A very high quality of entanglement is verified by various measurements, i.a. a tomographic state reconstruction is performed leading to a background corrected fidelity of $(99.45 \pm 0.06)\%$. Moreover, we describe an easy method for extending our source to arbitrarily high dimensions.

## 1. Introduction

On-chip integrated photonic circuits offer many advantages compared to bulk optical setups. Due to the nature of integration, circuits exhibit high phase stability while a great number of different optical devices can be combined in one circuit [1,2]. Therefore, integrated optics allows increasing complexity of optical circuits. This opens the door for more complex quantum optical systems. In particular, it is possible to realize a device called NxN multiport. This N input / N output device derives its potential from the fact that *any* N-dimensional unitary transformation (quNit operation) can be realized by a combination of qubit operations acting on different modes [3,4]. Therefore, depending on the internal parameters of the multiport (setting of the different qubit operations) any NxN unitary transformation can be realized. Using such a multiport would allow for a very general and flexible setup for different quantum optical experiments [4,5]. Important examples are higher-order Einstein-Podolsky-Rosen type perfect correlations between two entangled higher dimensional quantum systems [6], the generation and manipulation of entangled states [7,8], state discrimination [9], quantum key distribution [10], dense coding [11] and the problem of finding a complete set of mutually unbiased bases [12].

Beam splitters and phase shifters are standard optical elements ubiquitously used in integrated optics technology. Therefore, path-encoding is the natural way to implement N-dimensional photonic quantum states. Consequently, path-encoding quantum information on a chip has attracted wide interest in the community [1,2,13-17]. In order to fully use the advantage of integration, an entanglement source should be compatible with on-chip path-encoding, scalable in terms of its complexity when increasing the dimension N and ultimately allow integration as well. In this paper we report an experimental realization of a source meeting those requirements.

## 2. Principle of operation of an entangled quNit source

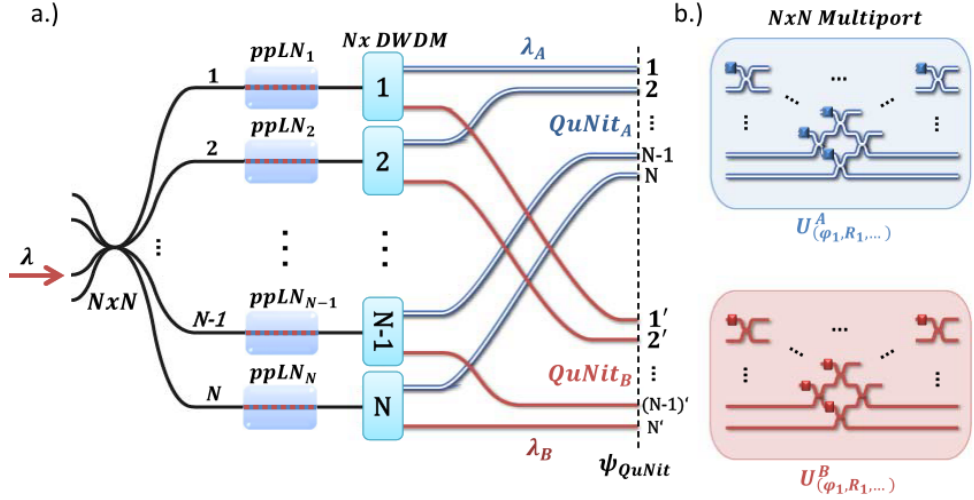

Fig. 1. Example of a setup for creating and manipulating path-entangled quNit pairs. a.) N non-linear crystal waveguides (ppLN) are used offering the feature of integration together with a higher down-conversion efficiency compared to bulk crystals [18,19]. All N crystals are coherently pumped by a common pump beam $\lambda$ split on a NxN beam splitter. With a certain probability a photon pair is created via type-I spontaneous parametric down-conversion (SPDC): $\lambda \rightarrow \lambda_A + \lambda_B$. Due to the small conversion probability, the possibility of multiple SPDC events occurring in one or more crystals at the same time is negligible. Therefore, coherent pumping of N crystals will result in a superposition of the SPDC event happening in one of the N ppLN crystals. In the following step the SPDC pairs $(\lambda_A, \lambda_B)$ in each mode (1,2,...,N) are separated by their wavelength using N dense wavelength division multiplexers (DWDM) into the two modes $1 \rightarrow (1_A, 1'_B), .... , N \rightarrow (N_A, N'_B)$. After regrouping the modes by their wavelength, a path-entangled two quNit state is obtained as given in Eq. (1). b.) Each photon then enters an NxN multiport, realized by a combination of phase shifters and beam splitters. By choosing the appropriate phase $(\varphi_i)$ and reflectivity $(R_i)$ settings any arbitrary N-dimensional unitary transformation can be realized [4]. Combined with single photon detection a projective measurement is finally realized (section 2.1).

The design of our path-entangled quNit source is depicted in Fig. 1(a). N non-linear crystal waveguides are coherently pumped by a common pump beam. This leads to a superposition of N down-conversion events happening in one of N crystals. The SPDC pairs then are separated by their wavelength using N dense wavelength division multiplexers (DWDM). Regrouping the modes by their wavelengths results in the following path-entangled two quNit state given in mode representation:

$$|\Psi> = \alpha_1 |1_A, 1'_B> + \alpha_2 e^{-i\varphi_1} |2_A, 2'_B> + ... + \alpha_N e^{-i\varphi_{N-1}} |N_A, N'_B>, \qquad (1)$$

where the phases $\varphi_i$ represent the total accumulated phases, while the amplitudes $\alpha_i$ are controlled by the splitting ratios of the *NxN* beam splitter. Note that in this scheme the photon pairs are separated by their wavelengths and therefore need to be non-degenerate. However, a degenerate path-entangled source can be obtained by replacing the type-I non-linear medium by a type-II SPDC process allowing separation by polarization.

As will be discussed in section 5, the important advantage of this particular design is the possibility to easily extend it to higher-dimensional quNits offering linear scalability in terms of the complexity of the source.

*2.1 Measuring path-encoded quNits*

In order to perform different measurements at the output an N input/ N output multiport device can be added (Fig. 1(b)) combining all N arms of each quNit. Depending on its internal phase and reflectivity setting the multiport is capable of realizing any unitary transformation [4]. Together with single photon detectors monitoring the output ports this is equivalent to a projection onto a specific state defined by the unitary transformation.

Setting the multiport to a specific balanced setting will lead to non-local perfect Einstein-Podolsky-Rosen (EPR) correlations between the two quNits [6]. More specific, N different higher order perfect EPR correlations between detectors on each quNit side are expected resulting from N different input phase relations of the quNits entering the multiport [6]. In principle, the two projective measurements on each quNit could be separated by arbitrary distance and still show the same perfectly correlated results. In Fig. 1(b) this is emphasized by the differently colored boxes around each projective measurement.

## 3. Experimental realization of an entangled two qubit source

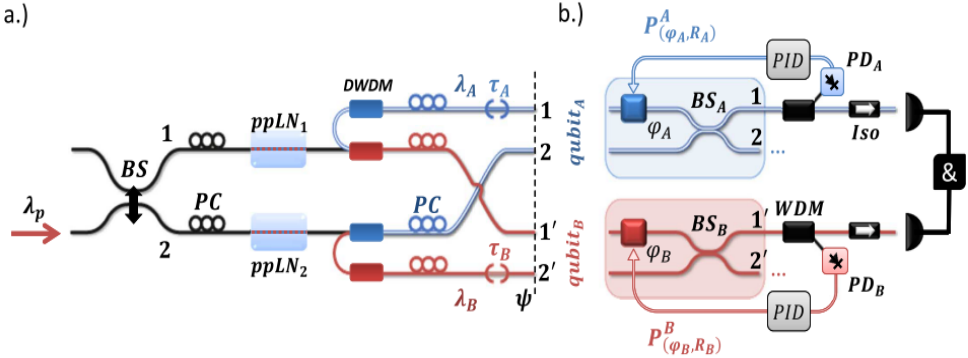

Fig. 2. The experimental setup for creating two path entangled qubits. a.) The pump beam λ is split by a variable beam splitter (BS) into the two modes 1 and 2. The splitting ratio is adjusted by changing the distance between the two fibers using a micrometer screw. Each mode enters a non-linear periodically poled Lithium Niobate waveguide (ppLN) creating photon pairs via spontaneous parametric down-conversion. Cascaded dense wavelength division multiplexers (DWDM) separate and spectrally filter the down-converted photon pairs. Modes 1 and 2 (1' and 2') define a path-encoded qubit. This leads to the two qubit path-entangled state. Delay lines ($\tau$) and polarization controller (PC) are used to adjust the arrival time and polarization of each mode. b.) 50/50 Beam splitters ($BS_A$, $BS_B$) and phases ($\varphi_A$, $\varphi_B$). Combined with single photon detection the projective measurement $|P(1/2,\varphi)><P(1/2,\varphi)|$ is realized (Eq. 2). Before entering the single photon detectors for coincidence detection (&) the signal and pump beams are separated using a WDM. For further separation an isolator (Iso) is added absorbing 775nm but passing 1550nm. After the WDM the separated pump beam is detected using standard photo diodes (PD). A PID controller uses this signal to stabilize the phase (sec. 3.3).

To demonstrate the experimental feasibility of our design we realize a source for two path-entangled qubits shown in Fig. 2(a). There have been a number of different entangled photon sources in the telecom band based on non-linear waveguides [18, 20] or photonic crystal fibers [21]. Their entangled degree of freedom was so far limited to time-bin or polarization entanglement. The concept of path-entanglement was first developed in [22,23] and generalized to higher-dimensional states in [3, 6]. It has been experimentally demonstrated for qubits [24] and extended to higher dimensional states [25]. Multiplexing different down-conversion sources can also be used to implement on-demand single-photon sources [26,27]. Here, the pump beam is demultiplexed into different spatial modes entering multiple non-linear crystals followed by separation of the photon pair by wavelength. All fibers are single mode around 1550nm, with the exception of the fibers before the non-linear waveguides, which are single mode for 780nm. All devices are standard off-the-shelf commercial devices used in the telecommunication industry.

A continuous-wave grating-stabilized laser at approximately 775nm wavelength is used as a pump beam. The splitting ratio is adjusted by a standard variable in-fiber beam splitter. Two 30mm long periodically-poled type-I MgO-doped Congruent Lithium Niobate waveguides (ppMgO:CLN) packaged and coupled to fibers are used for SPDC. We measured a down-conversion efficiency of approximately $10^{-5}$ to $10^{-6}$ per pump photon for the full output spectrum of around 5000GHz bandwidth. A series of two dense wavelength division multiplexers (DWDM) is used in each arm to separate and filter the photon pairs into the two standard 100GHz wide C-Band channels #32 (1551.721nm) and #36 (1548.515nm). Piezo fiber stretchers acting as phase shifters are used for phase control. Single-photon detectors based on InGaAs avalanche photo diodes (APDs) are used for coincidence detection. Before entering each single-photon detector the pump beam is filtered out using a combination of in-fiber isolators and wavelength division multiplexers (WDM).

After the non-linear waveguides all connections are spliced to avoid loss and back reflection of standard fiber connectors. The loss of the source from the point the single photons are created up to the state $\psi$ (Fig. 2(a)) is 1.9dB per arm. This originates from the intrinsic insertion loss of the different devices in each path and includes the estimated coupling loss of approximately 20% from the non-linear crystal waveguide to single-mode fiber. Note that the loss before the non-linear waveguides is not relevant. The projective measurement (Fig. 2(b)) introduces an additional loss of 0.2dB. Filtering and separating the pump light from the signal results in a loss of 0.8dB. This leads to a total loss of approximately 2.9dB. Additionally, our detectors feature around 10% detection efficiency contributing another 10dB per arm to the measured coincidence count rates. It is important to note that this loss is independent of the dimension of the encoded quNit.

*3.1 Measuring path-encoded qubits*

In order to perform different measurements at the output, a beam splitter is added combining the arms of each qubit (see Fig. 2(b)) followed by single-photon detection. This is equivalent to a projection onto the state

$$|P^A_{(\alpha,\varphi)}> = \sqrt{\alpha}|1_A> + \sqrt{1-\alpha}e^{-i\varphi_A}|2_A>, \qquad (2)$$

depending on the phase setting $\varphi_A$ and reflectivity $\alpha$ (and analogously for the second qubit B at paths 1' and 2'). Control of the two parameters, splitting ratio and phase, allows for any general unitary transformation in the qubit case. A particular measurement setting will lead to non-local perfect Einstein-Podolsky-Rosen (EPR) correlations between the two qubits [3,6].

*3.2 Indistinguishability*

In order to obtain a superposition of the two SPDC events on a beam splitter (Fig. 2(b)), the two events must be indistinguishable. Therefore, it is crucial that the outputs show the same physical properties (spectrum, arrival time, polarization). The central wavelengths are assured to be the same within $\pm 0.005nm$ by selecting the appropriate DWDM filters. The polarization is adjusted to be the same using polarization controller. The same arrival time of the photons of a pair is ensured by making the in-fiber optical path lengths the same within about $\pm 20 \mu m$ (to be compared to 100GHz bandwidth and therefore a coherence length on the order of 1mm). For this purpose optical in-fiber delay lines are used in one of the paths for each qubit to adjust the path difference. After adjustment the delays are locked and do not need to be adjusted again.

*3.3 Phase control*

Due to temperature fluctuations and external vibrations the phase difference of different fiber paths will change over time. Therefore, the phases are actively stabilized in the following way: light of the pump beam is transmitted through the whole setup accumulating a phase proportional to approximately twice the phase of the 1550nm photons. After interfering at the beam splitter (Fig. 2(b)) the 775nm light is coupled out by a WDM and detected. Depending on the phase changes in the fiber the detected 775nm intensity will vary allowing to determine the phase for the single photons at 1550nm. A PID controller uses this signal to stabilize and control the phase using the phase shifter setting $\varphi_A$ and $\varphi_B$.

The relation between the phases of the 775nm pump light and the 1550nm single photons varies slowly (~hrs) over time due to temperature drifts. Therefore, before each experimental run, the PID controller automatically characterizes the relation between the phase of the pump laser and the 1550nm single photons in a one minute characterization measurement. When reaching the limit of the piezo phase shifter during stabilization the corresponding $2\pi$ voltage is subtracted allowing in principle continuous operation.

**4. Results**

The source exhibits a high brightness and can reach detected coincidence rates of 1kHz at 100Ghz bandwidth and an input pump power of $250 \mu W$ into each non-linear crystal. However, in order to decrease both the dark count probability and saturation effects of the photon detectors as well as to lower the number of filter devices required to block the 775nm light, lower pump powers are used leading to a coincidence count rate (CC) of roughly $CC \approx 150 Hz$ and an accidental coincidence rate of $(1.47 \pm 0.15) Hz$. Thus, the coincidence-to-accidental-ratio (CAR) is around 100. The accidental coincidence rate is estimated by introducing a time delay in the coincidence detection system such that any detected coincidence signal cannot originate from a true signal pair. The origin of the accidental coincidences is a systematic effect of a finite coincidence window of around 2.5ns (much longer than the coherence time of the photons) due to the timing jitter of the detector together with a combination of the intrinsic detector dark count rate, the residual pump light and photons at the signal wavelength originating from fluorescence inside the non-linear crystal.

Different measurement settings are realized by combining the two modes of each qubit on a beam splitter (Fig. 2(b)). The phases $\varphi_A$ and $\varphi_B$ now represent different projection operations corresponding to a specific measurement setting (Eq. (2)). In the following three experiments are presented from which a measure for the quality of source can be obtained.

*4.1 Visibility of entanglement*

The entanglement visibility of the source is given by $V = (CC_{Max} - CC_{Min})/(CC_{Max} + CC_{Min})$. Ideally, for a maximally entangled state the visibility would be $V = 100\%$, its deviation is a direct measure of the achieved indistinguishability of the two SPDC events and a measure of the quality of the source. Setting the phase difference to $0$ and $\pi$ the maximum $CC_{Max}$ and minimum $CC_{Min}$ coincidence count rates are obtained leading to a visibility of

$$V = (95.6 \pm 0.4)\% \;,\; V_C = (97.3 \pm 0.5)\% \tag{3}$$

in the equal-superposition basis with $V_C$ including subtraction of accidental coincidences. The reduction of the visibility from the perfect 100% is most likely a result of a residual distinguishability. This can come from slightly different center wavelengths and spectral shapes of the DWDM filters, from different arrival times and polarizations, different arm losses and minimal errors in setting the pump frequency not exactly to half the center frequency of the two qubits as well as intrinsic differences of the two non-linear crystal waveguides. We remark, that the accidental corrected visibility cannot be distinguished from 100% in the computational basis. Here, the computational basis corresponds to a direct measurement of the state $\psi$ (Fig. 2(a)) with the setup for realizing projective measurements (Fig. 2(b)) removed.

*4.2 CHSH inequality*

To further stringently verify the quality of entanglement of our source a CHSH Bell-like inequality is tested [28]. Above a value of 2 the result cannot be explained by any local-realistic hidden variable model. Thus, the corresponding state must be non-separable and therefore entangled. For a maximally entangled state quantum theory predicts a maximum value of $2\sqrt{2}$. Therefore, the S value is a measure for the quality of the source. The phases $\varphi_A$ and $\varphi_B$ are set to 16 different settings corresponding to the different angles for a Bell measurement. An integration time of 10s is used for each measurement setting. Directly from the measured count rates without any correction the S value is obtained as

$$S_{CHSH} = (2.70 \pm 0.03). \tag{4}$$

This corresponds to a violation of the CHSH inequality by more than $20\sigma$. Compared to the ideal value $2\sqrt{2}$ the measured value shows a reduction to $(95.5 \pm 0.7)\%$. This is in excellent agreement with the non-perfect measured visibility of $V = (95.6 \pm 0.4)\%$ (Sec. 4.1).

*4.3 State tomography*

The quantum state $\psi$ is measured in three mutually unbiased bases. In the case of polarization entangled photons the three bases would be $(H,V), (+,-)$ and $(L,R)$. Here, the state $\psi$ (Fig. 2(a)) is first measured directly (first base). Using $\varphi_A, \varphi_B$ and $BS_{1/2}$ (Fig. 2(b)) projective measurements in the other two bases are realized. After subtracting the residual accidental coincidences from the data, using maximum likelihood estimation, the density matrix of the state is reconstructed (Fig. 3) [29]. From this the state fidelity $F$, as a measure of overlap between the measured and ideal state, and tangle, as a common entanglement measure, are obtained [29]. Compared to the ideal $\Psi^-$ state, the state fidelity $F$ and tangle $T$ are estimated to

$$F = (96.86 \pm 0.15)\% \text{ and } T = (87.9 \pm 0.6)\%. \tag{5}$$

With compensation for systematic accidental noise we obtain:

$$F_C = (99.44 \pm 0.06)\% \text{ and } T_C = (97.9 \pm 0.2)\%. \tag{6}$$

These values are the intrinsic fidelity and tangle of the source. This agrees well with the previous measurements of the quality of the state. The errors are estimated by a Monte Carlo simulation of the reconstruction analysis assuming Poisson noise in the measured counts.

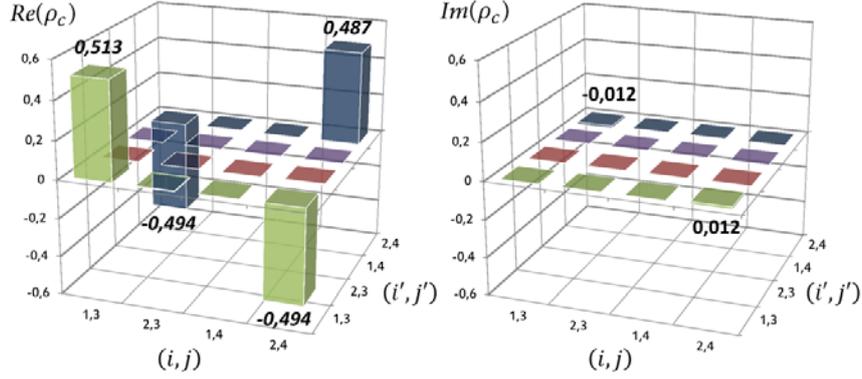

Fig. 3. Real and imaginary parts of the density matrix element $<i',j'|\rho_C|i,j>$ of the produced state $\psi$ reconstructed by maximum likelihood estimation after systematic accidental noise subtraction [29]. Unlabeled data corresponds to measured values smaller than $\text{Re}(\rho_C) < 0.005$ and $\text{Im}(\rho_C) < 0.0008$.

## 5. Extension to higher dimensions and outlook

The main advantage of our design is the possibility to extend it to higher-dimensional quNit systems. Following section 2 this can be achieved by splitting the pump beam coherently into N different paths followed by N non-linear waveguides resulting in a superposition of N down-conversion events. The separation by wavelength will lead to two path-entangled photons each defined in an N-dimensional Hilbert space. Thus, the entangled state is defined in an $N^2$-dimensional Hilbert space. For example, one can expand the entangled qubit source shown in Fig. 2(a) to an entangled ququart source by splitting the pump beam into four modes entering two additional non-linear crystals followed by separation by wavelength and resulting in:

$$|\Psi_{N=4}> = \alpha_1|1_A,1'_B> + \alpha_2 e^{-i\varphi_1}|2_A,2'_B> + \alpha_3 e^{-i\varphi_2}|3_A,3'_B> + \alpha_4 e^{-i\varphi_3}|4_A,4'_B>, \tag{7}$$

an entangled ququart state with amplitudes $\alpha, \beta, \gamma, \delta$ and phases $\varphi_i$. Note that the initial paths for the entangled qubits have not been altered and the performance of the qubit part is not influenced by increasing the dimension. For practical realizations, 1xN beam splitters integrated on-chip are commercially available at low cost with more than 100 outputs. Alternatively, cascaded variable in-fiber beam splitters could be used. Most commercially available non-linear crystal waveguides contain multiple parallel waveguides on one chip. Therefore, instead of using separate non-linear crystals, one crystal containing many waveguides could be used. Pig tailing and packaging this crystal to fiber arrays is a standard method and would further simplify the setup. Therefore, the number of components and thus the complexity of the setup scales linearly with the dimension N allowing a relatively easy access to higher dimensional quNits.

We have successfully demonstrated a novel, fully fiber integrated source for path-entangled photons. Our source is compatible with on-chip path-encoding of integrated photonic circuits. Its good scalability with respect to the dimensionality makes it an excellent design for complex experiments in higher dimensions. Due to the exceptionally good mode control of our all-integrated design, our source exhibits high brightness and very high fidelity of the produced entangled state. Moreover, there are no movable parts for alignment resulting in a high robustness and fully automated control. In the future it could be integrated on a single photonic chip.

Our experiment is a first step towards a general and flexible platform for different types of quantum optical experiments in higher-dimensional Hilbert spaces. Some example are the generation and manipulation of entangled states [7,8], state discrimination [9], or dense coding [11]. High-dimensional entangled states are an essential step for a deeper understanding of quantum information processing and foundations of quantum mechanics [6,12,30,31]. Additionally, quNits have been shown to provide a higher level of security of quantum information transfer [10] and higher robustness against decoherence [32]. Furthermore, from an applicational viewpoint the compatibility to telecom technology and fiber networks is another significant advantage of our design.

**Acknowledgement**
This work was supported by the ERC (Advanced Grant QIT4QAD, 227844), and the Austrian Science Fund FWF within the SFB F40 (FoQuS) and W1210-2 (CoQuS).